\documentclass[a4paper]{jpconf}
\usepackage[backend=biber,style=ieee]{biblatex}
\usepackage{graphicx}
\usepackage{xcolor}
\usepackage{hyperref}
\usepackage{indentfirst}
\usepackage[ruled,vlined]{algorithm2e}

\usepackage{caption}
\usepackage{subcaption}

\begin{document}
\title{Supernova Light Curves Approximation based on Neural Network Models}

\author{Mariia Demianenko$^{1,3}$, Ekaterina~Samorodova$^4$, Mikhail~Sysak$^3$, Aleksandr~Shiriaev$^{5}$, Konstantin~Malanchev$^{6,2}$, Denis~Derkach$^{1}$,  Mikhail~Hushchyn$^1$}

\address{$^1$HSE University, 20 Myasnitskaya Ulitsa, Moscow 101000, Russia;}
\address{$^2$Lomonosov Moscow State University, Sternberg Astronomical Institute, Universitetsky pr. 13, Moscow 119234, Russia;}
\address{$^3$Moscow Institute of Physics and Technology, Institutskii Pereulok 9, Dolgoprudny, Moscow Region 141700, Russia;}
\address{$^4$Lomonosov Moscow State University, Faculty of Space Research, Leninskiye Gory 1, bld. 52, Moscow 119234, Russia;}
\address{$^5$Moscow Polytechnic University, Tverskaya street, 11, Moscow 125993, Russia;}
\address{$^6$Department of Astronomy, University of Illinois at Urbana-Champaign, 1002 West Green Street, Urbana, IL 61801, USA.}

\ead{mhushchyn@hse.ru}

\begin{abstract}
Photometric data-driven classification of supernovae becomes a challenge due to the appearance of real-time processing of big data in astronomy. Recent studies have demonstrated the superior quality of solutions based on various machine learning models. These models learn to classify supernova types using their light curves as inputs. Preprocessing these curves is a crucial step that significantly affects the final quality. In this talk, we study the application of multilayer perceptron (MLP), bayesian neural network (BNN), and normalizing flows (NF) to approximate observations for a single light curve. We use these approximations as inputs for supernovae classification models and demonstrate that the proposed methods outperform the state-of-the-art based on Gaussian processes applying to the Zwicky Transient Facility Bright Transient Survey light curves. MLP demonstrates similar quality as Gaussian processes and speed increase. Normalizing Flows exceeds Gaussian processes in terms of approximation quality as well.
\end{abstract}

\section{Introduction}

Currently, a number of transients discovered by photometric surveys is increasing rapidly. 
Thus, automated light-curve processing is crucial for various tasks: from deriving phenomenological parameters of the object, like peak time and magnitude, to machine-learning-based photometric classification.
Usually, light curves have a different cadence in different photometric passbands, complicating the machine-learning models and increasing their computational time.
Therefore, fast and accurate approximation of time series is highly important in order to obtain homogeneous data. Currently, the SVM algorithm~\cite{doi:10.1142/S0218271810017160}, decision trees and clustering algorithms~\cite{baron2019machine}, as well as probabilistic approaches such as regression of Gaussian processes~\cite{10.1093/mnras/stx2109} are used to approximate astronomical time series. In this paper, we consider neural network models for the augmentation of observations with the same time step inside the light curve. A recent study shows a deterioration in quality metrics when applying machine learning models to real data compared to simulation data ~\cite{Dobryakov2021}, we test our models on a data catalog of real nature.

In this paper, we report the tests of such neural models as multilayer perceptron (MLP), Bayesian neural network (BNN), normalizing flows (NF), as they proved to be much better than other tested models (SVM regression, Radial basis function network, FE, XGBoost, CatBoost). We used regression quality metrics to solve the problem of the light curve approximation and indirect physical metrics: binary classification for supernovae Ia and the rest, estimation of the bolometric peak.


\begin{figure}
     \centering
     \begin{subfigure}[b]{0.49\textwidth}
         \centering
         \includegraphics[width=\textwidth]{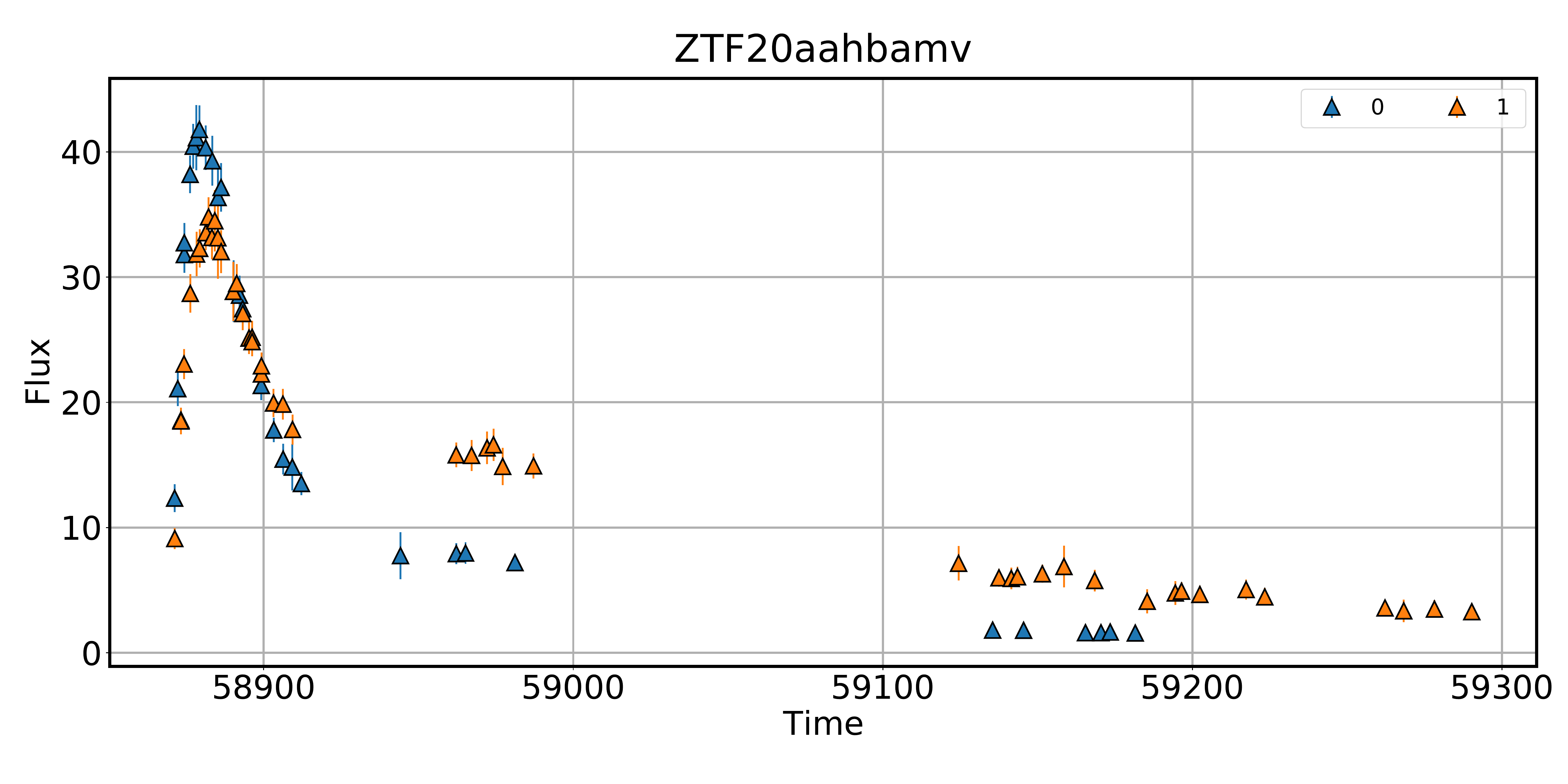}
         \caption{No approximation.}
         \label{fig:y equals x}
     \end{subfigure}
     \hfill
     \begin{subfigure}[b]{0.49\textwidth}
         \centering
         \includegraphics[width=\textwidth]{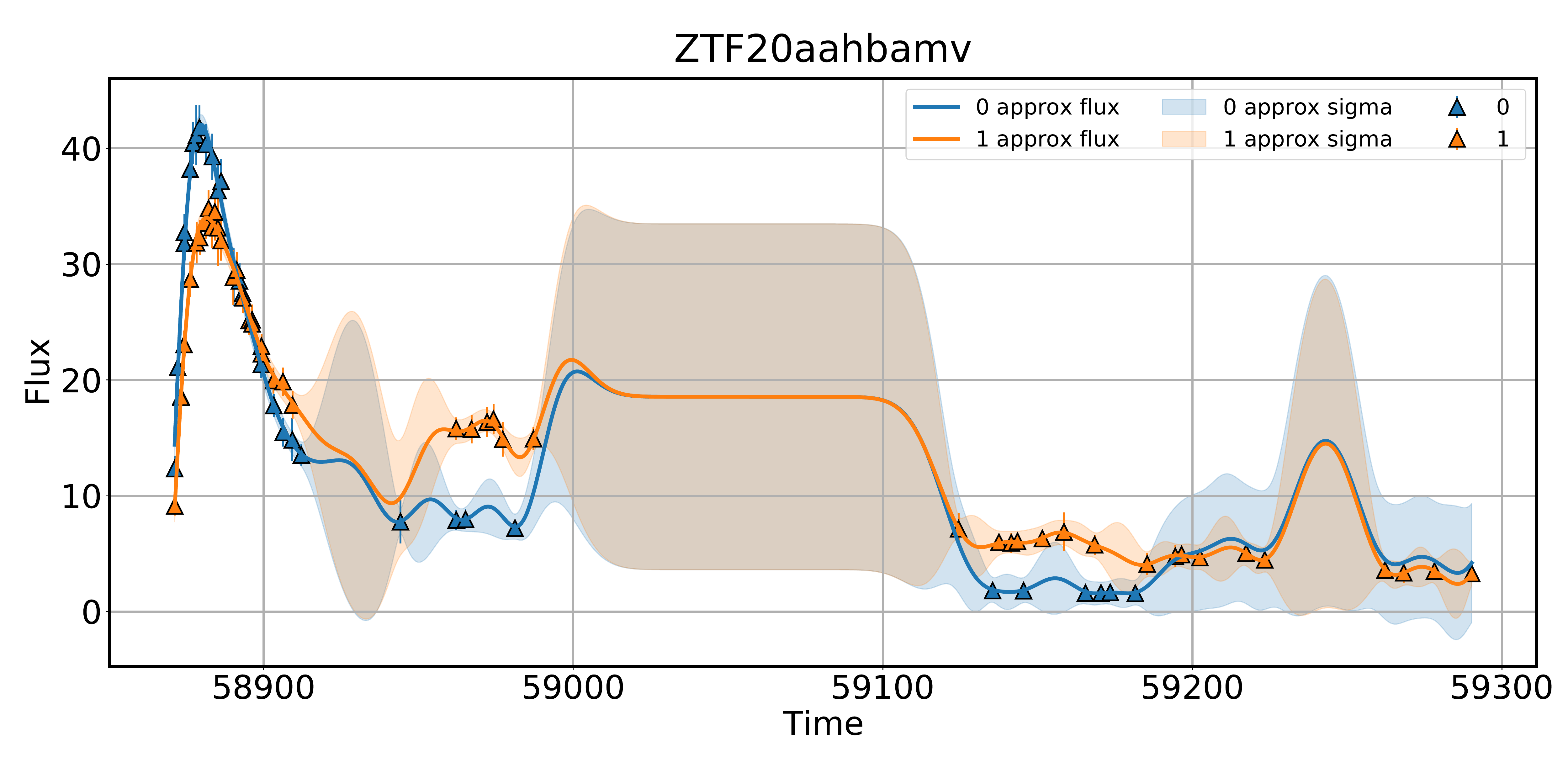}
         \caption{GP approximation.}
         \label{fig:three sin x}
     \end{subfigure}
     \hfill
     \begin{subfigure}[b]{0.49\textwidth}
         \centering
         \includegraphics[width=\textwidth]{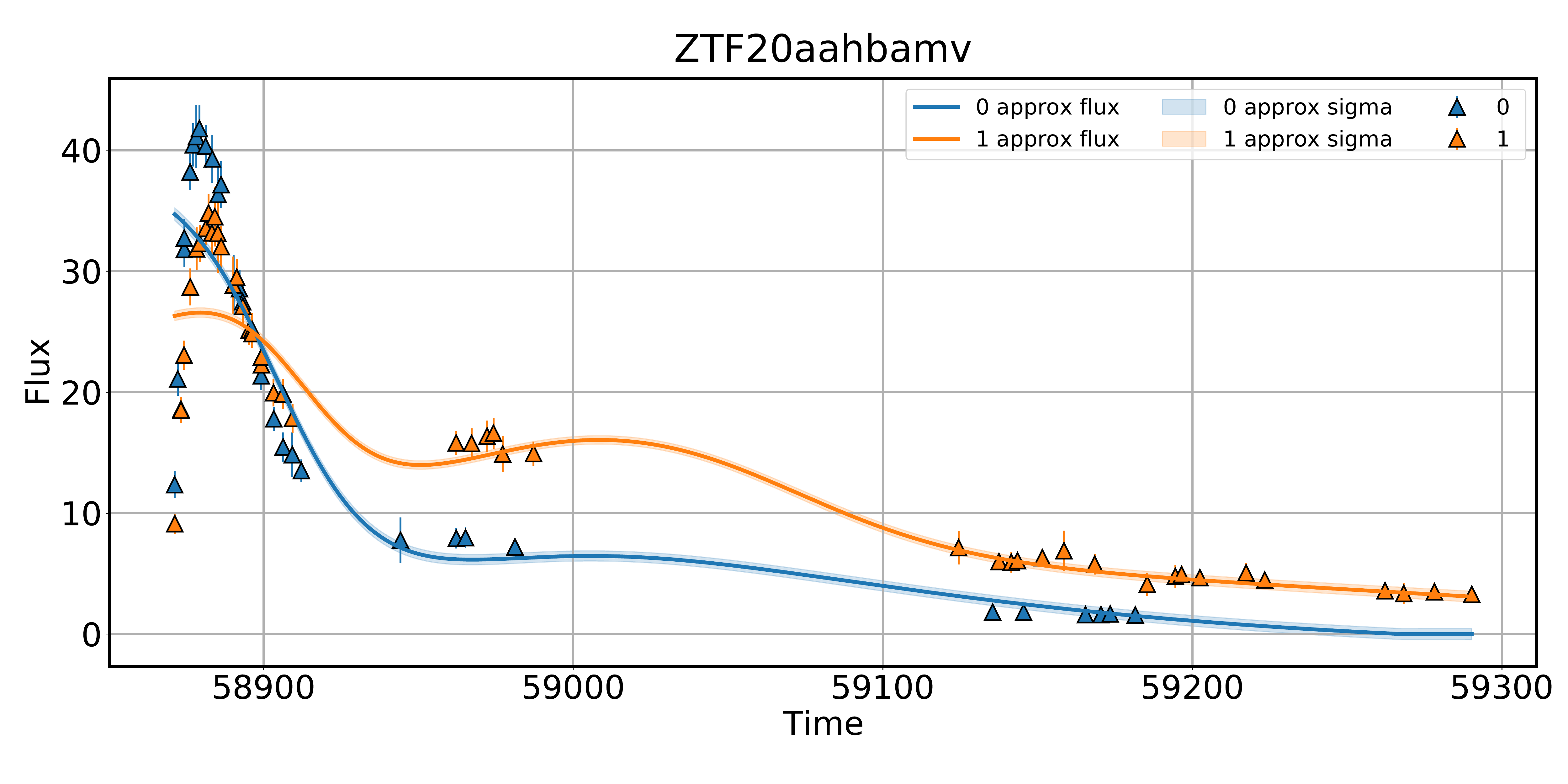}
         \caption{BNN approximation.}
         \label{fig:three sin x}
     \end{subfigure}
     \hfill
     \begin{subfigure}[b]{0.49\textwidth}
         \centering
         \includegraphics[width=\textwidth]{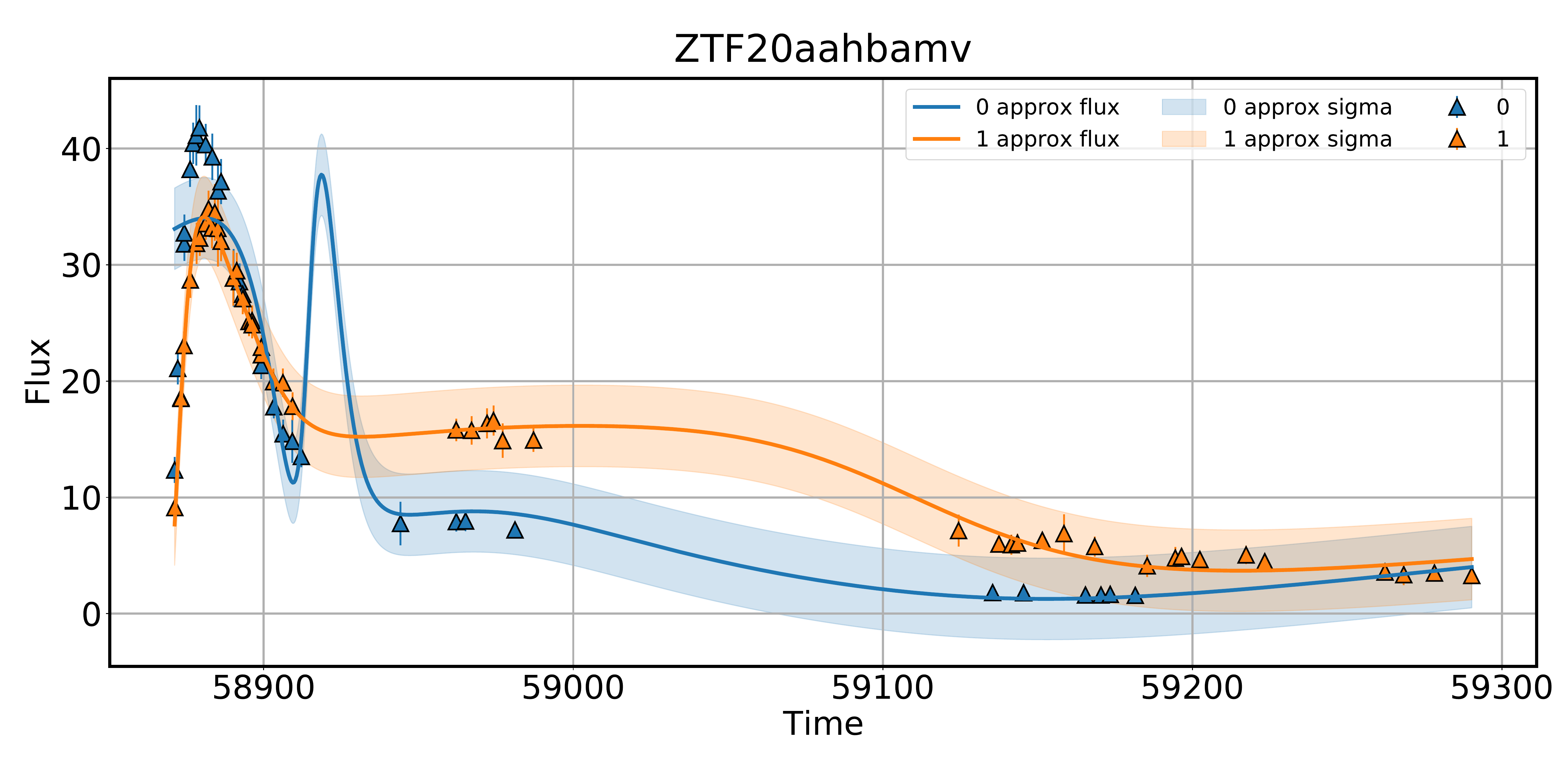}
         \caption{MLP (scikit-learn) approximation.}
         \label{fig:three sin x}
     \end{subfigure}
     \hfill
     \begin{subfigure}[b]{0.49\textwidth}
         \centering
         \includegraphics[width=\textwidth]{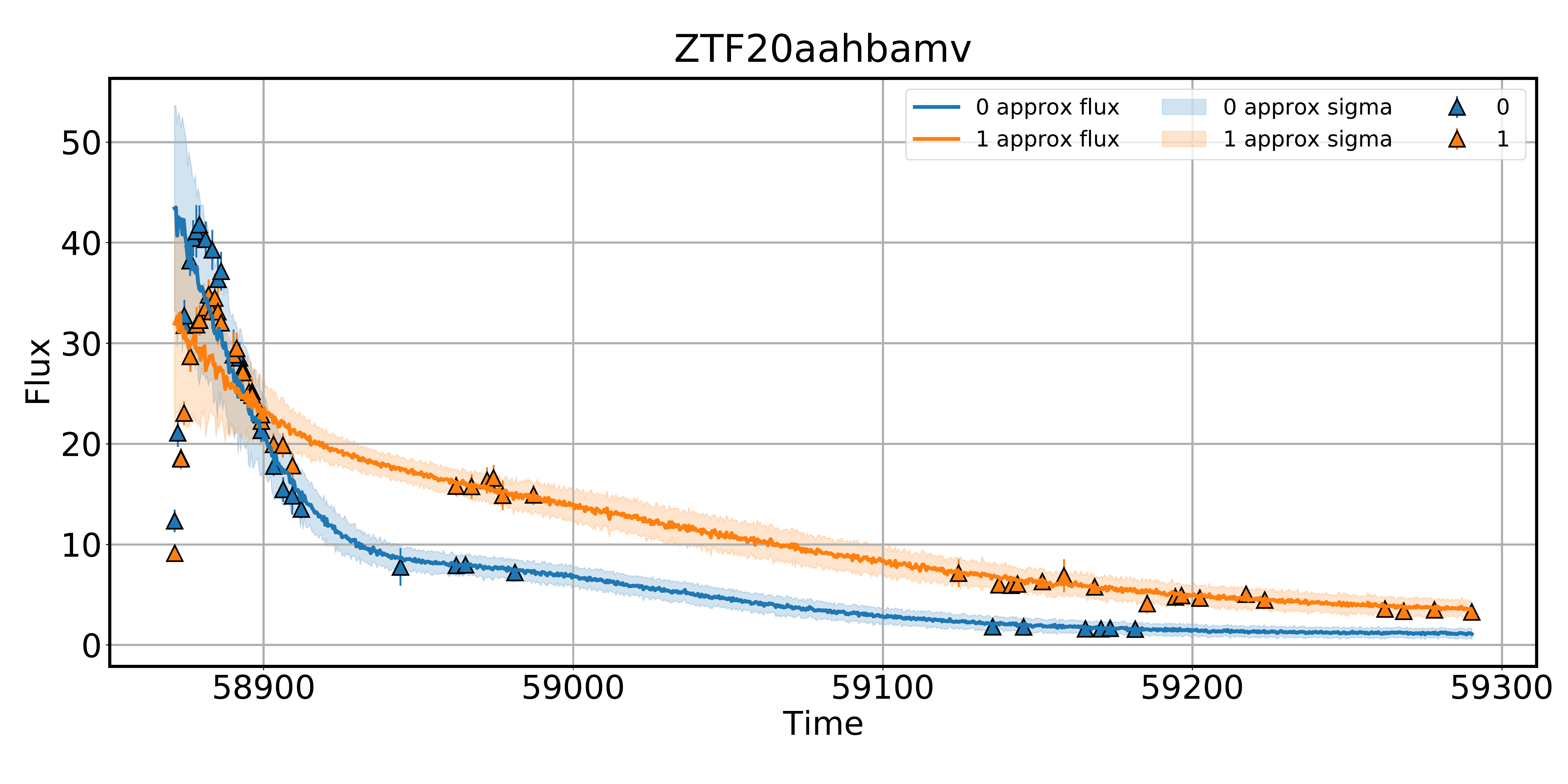}
         \caption{NF   approximation.}
         \label{fig:three sin x}
     \end{subfigure}
     \hfill
     \begin{subfigure}[b]{0.49\textwidth}
         \centering
         \includegraphics[width=\textwidth]{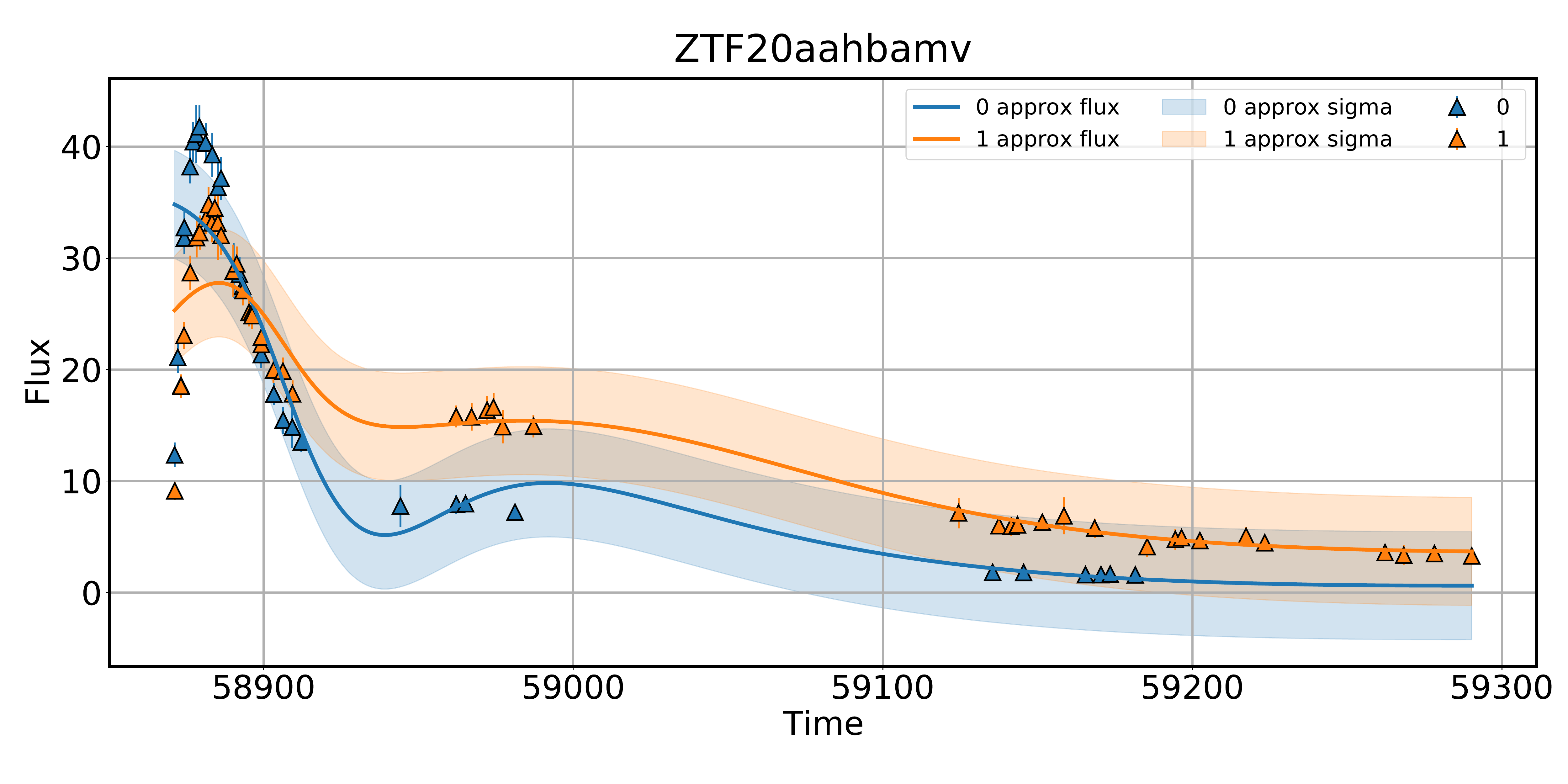}
         \caption{MLP (pytorch) approximation.}
         \label{fig:three sin x}
     \end{subfigure}
        \caption{Light curve of ZTF20aahbamv (supernova type II) before and after approximation. The points represents measurements in the corresponding passband. The shaded area represents $\sigma$ uncertainty band for the light curve approximation.}
        \label{approx_2}
\end{figure}

\section{Data}
The light curve $y(t, p)$ is a time series of the radiation flux, which presents observations for time points $t$ and passbands $p$ with different time cadences. For example, such a time series for one object is shown in Fig. \ref{approx_2}. The real The Zwicky Transient Facility (ZTF) Bright Transient Survey dataset of transient light curves, that is, astronomical events of extreme flux variation contains more than 3000 objects brighter than magnitude 19. However, in our work, 2493 light curves were used, which have at least a total of 10 observations in 2 photometric passbands $\{r, g\}$ for each object.

\section{Problem statement}
In approximation problem, the feature matrix $x_i = (t_i, p_i)$ consists of time moments $t_i$ and photometric passband $p_i$ for each observation $x_i$ in the light curve $y(t, p)$ of astronomical event. Accordingly, each observation $x_i$ have $y(x_i)$ value. Further, as in any regression problem, it is required to train a model to predict the value of an observation $y(x_i)= f(x_i) + \varepsilon_y(x_i)$ and estimate the error $\sigma(x_i)$ of such a value. The model is fitted by minimizing the Mean Squared Error (MSE) loss function $\sum(y(x_i) - f(x_i))^2 $. After training the model, augmented observation values are generated with a fixed uniform time step in each passband $p$. The time grid can contain as many steps as the user needs for a specific task. Thus, a two-dimensional array of sets of values for each observation $x_i$ is created at the approximation output. Each set contains values for all passbands $p$. This output of the model can be used for further classification tasks or evaluation of the bolometric peak. 

The result of approximation using a probabilistic regression model of Gaussian processes is illustrated in Fig.~\ref{approx_2}. A peculiarity of Gaussian processes is zero error at the learning points and a large error in the intervals where there are large time gaps between observations. \textbf{Gaussian processes (GP)} are chosen here as an example since they are very often used~\cite[see, e.g.][]{10.3847/1538-3881/ab5182} for problems of light curves approximation and were considered the best solution.

The task of this study is to improve the existing state-of-the-art approach. We use various neural network models, optimizing their architecture and hyperparameters. As a result, we settle on the three best models.

\section{Models}
We use a \textbf{multilayer perceptron (MLP)} from the \textit{scikit-learn}~\cite{scikit-learn} realization with 2 hidden layers with 20 and 10 neurons, {\tt tanh} activation function with {\tt LBFGS} optimizer. Such an optimizer proved to be the most efficient in terms of the convergence time of the solution. The model does not directly predict the error of the radiation flux value. However, we estimate it as the standard deviation of the model values. The training takes place using observation belonging to one light curve. Approximately 50 percent of the points are taken for training and 50 percent for testing the model.

A \textbf{Bayesian neural network (BNN)} with 2 linear layers of 15 and 7 neurons, respectively, and weight priors in the form of standard normal distributions $\mathcal{N}(0, 0.1)$ is also investigated. In such a model, the {\tt Adam} optimizer and the {\tt tanh} activation function are used. This model predicts radiation flux uncertainties as the final variance of the distribution of weights.

\textbf{Normalizing flows (NF)} with 8 Real-NVP transformations, where 2 simple fully connected neural networks are used in each transformation, is also considered. In this model, radiation flux errors are also predicted by construction.
The code of the neural network approximation library for python with comments and examples of use and plotting can be studied at the \textit{Fulu}\footnote{https://github.com/HSE-LAMBDA/fulu} repository.

\section{Results}
Tests for regression and classification were provided on AMD Ryzen 7 4700U laptop with 8 CPU and 16 GB RAM.
The quality of the approximated values is compared at the points deferred for the test in each curve based on the classical regression metrics Root-Mean-Square Error (RMSE), Mean Squared Error (MSE), Relative Absolute Error(RAE), Relative Standard Error (RSE), and Mean Absolute Percentage Error (MAPE) in Table~\ref{tab:regression_metrics}. Among these metrics, MAPE is the most significant since there may be different amplitudes of the radiation flux drop in different light curves, and the metrics are averaged as a result over all the light curves in the dataset. The specified time is the approximate processor time for all 2493 curves, taking into account each model's training and testing time. As expected, MLP with the Limited-memory Broyden–Fletcher–Goldfarb–Shanno (LBFGS) optimizer has the fastest data processing time. Nonetheless, if one requires the highest quality of the approximation, NF is the most optimal algorithm.

\begin{table}[h]
    \centering
    \def\arraystretch{1.2}
    \begin{tabular}{|c|cccccc|} \hline
         Model & RMSE & MAE & RSE & RAE & MAPE & Time \\ \hline
         GP    & $2.9 \pm 0.2$ & $2.1 \pm 0.1$ & $0.47 \pm 0.01$ & $0.43 \pm 0.01$ & $19.2 \pm 0.5$ & $02$:$38$ \\ \hline
         MLP (sklearn)  & $4.6 \pm 0.5$ & $3.0 \pm 0.3$ & $0.66 \pm 0.01$ & $0.57 \pm 0.01$ & $21.6 \pm 0.5$ & $00$:$43$ \\ \hline
         BNN & $4.2 \pm 0.3$ & $2.9 \pm 0.2$ & $0.59 \pm 0.01$ & $0.52 \pm 0.01$ & $19.8 \pm 0.3$ & $37$:$12$ \\ \hline
         NF & $4.0 \pm 0.3$ & $2.5 \pm 0.1$ & $0.53 \pm 0.01$ & $0.47 \pm 0.01$ & $17.5 \pm 0.3$ & $6$:$02$:$51$ \\ \hline 
    \end{tabular}
    \caption{Regression metrics for approximation models}
    \label{tab:regression_metrics}
\end{table}

\begin{figure}
     \centering
     \begin{subfigure}[b]{\textwidth}
         \centering
         \includegraphics[width=\textwidth]{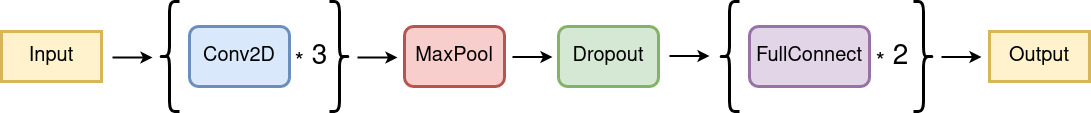}
         \label{fig}
     \end{subfigure}
     \caption{Binary classifier scheme.}
        \label{cnn}
\end{figure}
In addition to regression metrics, the metric of further classification of objects by their light curves is used. Tab.~\ref{tab:classification_metrics} demonstrate the results of binary classification on Supernovae (SN) type Ia and all other types. For the classification problem, a simple convolutional neural network (CNN) is used, which takes as input light curves processed by one of the algorithms above. The scheme of the classifier is shown in Fig.~\ref{cnn}. In this case, the most significant ones are the metrics Area Under Curve Receiver Operating Characteristic (AUC-ROC), Area Under Curve Precision-Recall (AUC-PR), since in our dataset, the classes are not balanced, and the light curves of SN Ia are more than 70$\%$. NF and BNN are the leaders in these metrics and beat the GP solution. Nevertheless, MLP is still the leader in the fastest good solution, differing slightly in quality compared to GP.

\begin{table}[h]
    \centering
    \def\arraystretch{1.2}
    \begin{tabular}{|c|cccc|} \hline
         Model & AUC-ROC & AUC-PR & Accuracy & Complexity \\ \hline
         GP   & $0.7603 \pm 0.0002$ & $0.8516 \pm 0.0002$ & $0.7916 \pm 0.0001$ & $O(N^3)$ \\ \hline
         MLP (sklearn) & $0.7392 \pm 0.0002$ & $0.8447 \pm 0.0002$ & $0.7497 \pm 0.0001$ & $O(N)$ \\ \hline
         BNN & $0.7933 \pm 0.0002$ & $0.8682 \pm 0.0002$ & $0.7837 \pm 0.0001$ & $O(N)$ \\ \hline
         NF & $0.8456 \pm 0.0001$ & $0.8956 \pm 0.0001$ & $0.7722 \pm 0.0001$ & $O(N)$ \\ \hline
    \end{tabular}
    \caption{Classification metrics for approximation models}
    \label{tab:classification_metrics}
\end{table}

\section{Conclusion}

This work shows the results of neural network approximation models on a real data sample using regression metrics and metrics to solve the classification problem on approximated curves. Multi-layer perceptron demonstrates the best approximation quality and speed ratio. Normalizing flows is the most accurate algorithm but significantly increases the operating time compared to GP. Thus, neural network models are better than classical approaches, now state-of-the-art solutions.

\section*{Acknowledgement}
KM work on data preparation is supported by the RFBR and CNRS according to the research project №21-52-15024. DD, MH, and MD are supported by the Academic Fund Program at the HSE University in 2022 (grant №22-00-025) in designing, constructing, and testing data augmentation techniques.


\printbibliography

\end{document}